\documentclass[a4paper]{jpconf}
\usepackage{graphicx}
\begin{document}
\title{Confinement and electron correlation effects in photoionization of
atoms in endohedral anions: Ne@C$_{60}^{z-}$}

\author{V K Dolmatov, G T Craven and D Keating}

\address{Department of Physics and Earth Science, University of North Alabama, Florence, AL 35632, USA}

\ead{vkdolmatov@una.edu}

\begin{abstract}
Trends in resonances, termed confinement resonances, in photoionization
of atoms $A$ in endohedral fullerene anions $A$@C$_{60}^{z-}$ are
theoretically studied and exemplified by the photoionization of Ne in
Ne@C$_{60}^{z-}$. Remarkably, above a particular $nl$ ionization
threshold of Ne in \textit{neutral} Ne@C$_{60}$ ($I_{nl}^{z=0}$),
confinement resonances in corresponding partial photoionization cross
sections $\sigma_{nl}$ of Ne in any charged Ne@C$_{60}^{z-}$ remain
almost intact by a charge $z$ on the carbon cage, as a general
phenomenon. At lower photon energies, $\omega < I_{nl}^{z=0}$, the
corresponding photoionization cross sections develop additional, strong,
$z$-dependent resonances, termed Coulomb confinement resonances, as a
general occurrence. Furthermore, near the innermost ${\rm 1s}$
ionization threshold, the ${\rm 2p}$ photoionization cross section
$\sigma_{\rm 2p}$ of the outermost ${\rm 2p}$ subshell of thus confined
Ne is found to inherit the confinement resonance structure of the
 ${\rm 1s}$ photoionization spectrum, via interchannel
coupling. As a result, new confinement resonances emerge in the ${\rm
2p}$ photoionization cross section of the confined Ne atom at
photoelectron energies which exceed the ${\rm 2p}$ threshold by about a
thousand eV, i.e., far above where conventional wisdom said they would
exist. Thus, the general possibility for confinement resonances to
resurrect in photoionization spectra of encapsulated atoms far above
thresholds is revealed, as an interesting novel general phenomenon.

\end{abstract}

\section{Introduction}
Endohedral fullerenes $A$@C$_{60}$, where the atom $A$ is encapsulated inside the hollow interior of the carbon cage C$_{60}$,
are of the highest
interest and importance both to the basic and applied sciences and technologies, as new modern building blocks of materials and devices with unique
properties. Therefore, they have attracted much attention of many investigators in recent years. In particular,
the photoionization, as a basic phenomenon in nature, of atoms $A$ in carbon fullerenes $A$@C$_{60}$ has become a topical research subject both for
theorists for some years now (see a review paper~\cite{DolmAQC09} as well some latest works~\cite{PranawaXe09,Himadri09,GiantCRs08} on the subject and references therein)
and, since only very recently, experimentalists~\cite{MuellerJPCS07,MuellerPRL08}.  Among the performed photoionization studies of thus
confined atoms, only works~\cite{DolmAQC09,Ne@C60z} have provided the initial understanding of how the photoionization spectrum of an atom
could be modified by the environment of a negatively charged carbon cage C$_{60}^{z-}$. The understanding was exemplified by
trends in photoionization of the innermost ${\rm 1s}$ subshell of Ne in Ne@C$_{60}^{z-}$ with various $z$'s.
However, how these and other possible trends might show up, as general phenomena, in spectra of
intermediate and outer subshells of atoms $A$ in $A$@C$_{60}^{z-}$ has remained unstudied. The present paper  expands further the investigation started in~\cite{Ne@C60z}
to the intermediate ${\rm 2s}$ and outermost ${\rm 2p}$ subshells of the confined Ne
with the aim to reveal which new aspects of these spectra of endohedral anions are most interesting, as general phenomena.

\section{Brief description of theoretical concepts}
Following previous works, see, e.g.,
 \cite{DolmAQC09,PranawaXe09,Ne@C60z} and references therein,
the neutral C$_{60}$ cage
is modelled by a short-range attractive
spherical potential $V_{\rm c}(r)$ of inner radius $r_{0}=5.8$ a.u., depth
$U_{0}=-8.2$ eV and finite thickness
$\Delta= 1.9$ a.u.
\begin{eqnarray}
 V_{\rm c}(r)=\left\{\matrix {
U_{0}, & \mbox{if $r_{0} \le r \le r_{0}+\Delta$} \nonumber \\
0 & \mbox{otherwise.} } \right.
\label{eqVc}
\end{eqnarray}
A neutral endohedral fullerene $A$@C$_{60}$ is formed by placing the
atom $A$ at the center of the cage.
For small sized, compact
atoms $A$ there is no charge transfer to the cage, so that the confined atom
$A$ retains the general structure of the free atom $A$.
Alternatively~\cite{DolmAQC09,Ne@C60z}, an endohedral anion $A$@C$_{60}^{z-}$ is modelled by
 the sum total of the potential $V_{\rm c}$ and the Coulomb potential $V_{z}(r)$ of an excessive negative charge
on the cage C$_{60}$. Assuming that the
charge $z$ is uniformly distributed over the entire outer surface of C$_{60}$,
\begin{eqnarray}
V_{z}(r)=\left\{\matrix {
\frac{z}{r_{0}+\Delta}, & \mbox{if $0 \le r \le r_{0}+\Delta$} \nonumber \\
\frac{z}{r} & \mbox{otherwise.} } \right.
\label{eqVq}
\end{eqnarray}
Next, the sum total of these two potentials is added to nonrelativistic Hartree-Fock (HF) equations for a
free atom.  Solutions of these
new HF equations, i.e., electronic energies and wavefunctions of the confined $A$ atom,
are used in well-known expressions for $nl$
photoionization amplitudes, angle-differential and/or
angle-integrated $nl$ photoionization cross sections,
\textit{etc.}, for free atoms; see~\cite{Amusia-book} for the latter.  To account for
interchannel coupling in the photoionization of a confined atom, the \textit{random phase approximation
with exchange} (RPAE) \cite{Amusia-book} is utilized to meet the aim. This is because RPAE,
which uses a HF approximation as the zero-order approximation,
 has proven to be a very reliable methodology over the years. Accordingly, for the sake of
``theoretical'' consistency, HF values of ionization thresholds of free and confined Ne atoms are used in the present study.

\section{Results and discussion}

\subsection{Conventional confinement resonances: irrelevance of a charged state of C$_{60}^{z-}$}

\begin{figure}[h]
\begin{center}
\includegraphics[width=9.5cm]{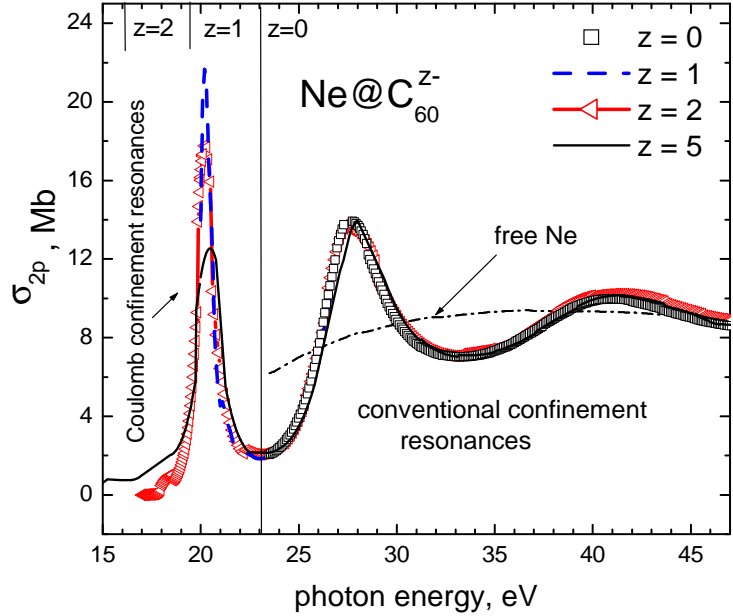}
\caption{
 RPAE calculated data, at an intra-shell interchannel coupling approximation level, for the $\sigma_{\rm 2p}$ photoionization cross section  of Ne in Ne@C$_{60}^{z-}$
 for $z=0$, $1$, $2$, $5$ as well as for free Ne, as marked. Vertical lines with marks $z=0$, $z=1$ and $z=2$, show
positions of the Ne ${\rm 2p}$ thresholds $I_{\rm 2p}^{z=0} \approx 23.1$, $I_{\rm 2p}^{z=1} \approx 19.6$,
 $I_{\rm 2p}^{z=2} \approx 16.1$ and $I_{\rm 2p}^{z=5} \approx 5.5$ eV
 in corresponding Ne@C$_{60}^{z-}$.
}
\label{Fig1}
\end{center}
\end{figure}

\begin{figure}[h]
\begin{center}
\includegraphics[width=8cm]{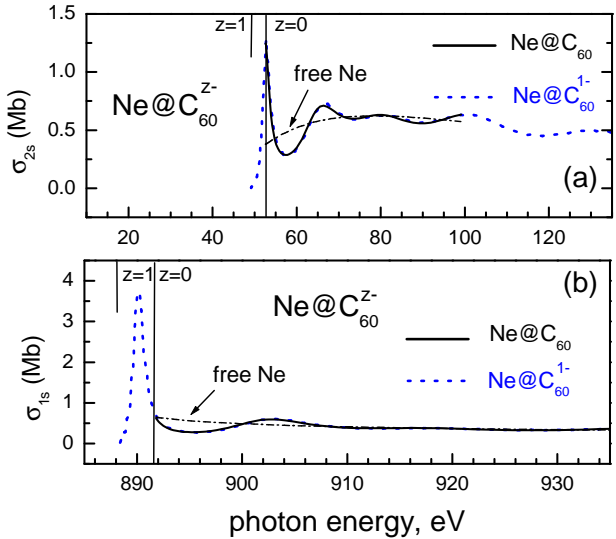}\hspace{2pc}%
\begin{minipage}[b]{16pc}\caption{\label{Fig2}
(a): RPAE calculated data, at an intra-shell interchannel coupling approximation level, for the $\sigma_{\rm 2s}$ photoionization cross section  of Ne in Ne@C$_{60}^{z-}$
 for $z=0$ and $1$ as well as for free Ne, as marked. Vertical lines with marks $z=0$ and $z=1$ show
positions of Ne $I_{\rm 2s}^{z=0} \approx 52.5$ eV and $I_{\rm 2s}^{z=1} \approx 49$ eV.
(b): The same as in (a) but for the Ne ${\rm 1s}$ photoionization ($I_{\rm 1s}^{z=0} \approx 892$, $I_{\rm 1s}^{z=1} \approx 888$ eV).
}
\end{minipage}
\end{center}
\end{figure}

\begin{figure}[h]
\begin{center}
\includegraphics[width=8cm]{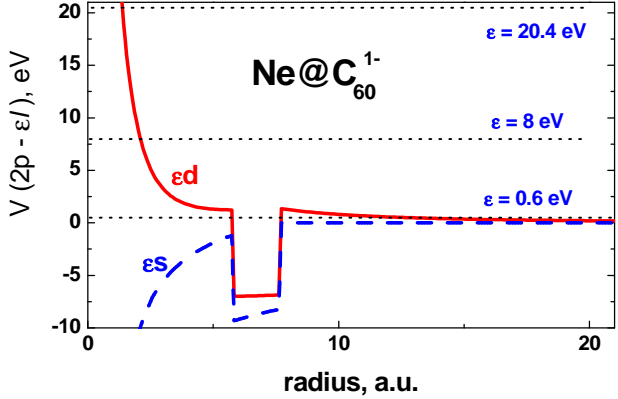}\hspace{2pc}%
\begin{minipage}[b] {16pc} \caption{\label{Fig3}
 A direct (Hartree) part of the potential $\rm{V(2p}$ $\rightarrow$ $\epsilon l)$ ``seen'' by $\epsilon\rm{s}$ and $\epsilon\rm{p}$ photoelectrons in the $\rm 2p$ $\rightarrow$ $\epsilon\rm{s}$ and
  $\rm 2p$ $\rightarrow$ $\epsilon\rm{d}$ photoionization transitions in the Ne@C$_{60}^{1-}$ atom, as marked. Horizontal lines show the photoelectron
  energies $\epsilon \approx 0.6$, $8$ and $20.4$ eV
 corresponding to
 maxima of the resonances in the Ne@C$_{60}^{1-}$  photoionization cross section $\sigma_{2p}$ (see figure~\ref{Fig1})
 at about $20$, $28$ and $42$ eV of the photon energy, respectively.
}
\end{minipage}
\end{center}
\end{figure}

RPAE calculated photoionization cross sections $\sigma_{\rm 2p}$ of Ne in variously $z$-charged Ne@C$_{60}^{z-}$ are displayed in figure~\ref{Fig1}.
Corresponding photoionization cross sections
$\sigma_{\rm 2s}$ and $\sigma_{\rm 1s}$ are depicted in figure \ref{Fig2}.
In these calculations, interchannel coupling in the atom was accounted for at an intra-shell approximation level, as the first step in the study.
One can see that, above the ${\rm 2p}$ threshold $I_{\rm 2p}^{z=0}\approx 23$ eV of Ne in \textit{neutral} Ne@C$_{60}^{z=0}$ (i.e., to the right of a vertical line marked with $z=0$),
all $\sigma_{\rm 2p}$'s oscillate about the ${\rm 2p}$ photoionization cross section
of free Ne, as do  $\sigma_{\rm 2s}$ above $I_{\rm 2s}^{z=0} \approx 53$ eV and $\sigma_{\rm 1s}$ above $I_{\rm 1s}^{z=0} \approx 892$ eV, respectively.
The oscillations are due to the interference between
the outgoing $nl$ photoelectron wave and those scattered off the confining potential of C$_{60}^{z-}$. When
the constructive interference occurs, there emerge maxima, i.e., resonances in $\sigma_{nl}$'s, termed confinement resonance~\cite{DolmAQC09,PranawaXe09,GiantCRs08,Ne@C60z}.
Furthermore, one can see that
confinement resonances  in \textit{all} $\sigma_{nl}$'s are nearly $z$ independent above related $I_{nl}^{z=0}$ thresholds (previously~\cite{Ne@C60z}, the same
was noted in the $\rm 1s$ spectrum of Ne in Ne@C$_{60}^{z-}$).
This is because the energy $\epsilon$ of an outgoing $nl$ photoelectron
in any charged Ne@C$_{60}^{z-}$ exceeds by far the Coulomb potential barrier of the charged carbon cage, at $\omega \ge I_{nl}^{z=0}$.
This is clearly seen in figure~\ref{Fig3} where direct (Hartree) parts of the potentials ``seen'' by the $\epsilon$p and $\epsilon$s photoelectrons (due to
the $\rm 2p$ $\rightarrow$ $\epsilon$p, $\epsilon$s transitions) are displayed. Hence, the presence of
the Coulomb potential barrier is inconsequential for the outgoing $nl$ photoelectrons at photon energies $\omega \ge I_{nl}^{z=0}$. Correspondingly,
at $\omega \ge I_{nl}^{z=0}$, confinement resonances in any $nl$ photoionization spectrum of the encapsulated atom will
chiefly be
governed by details of the confining potential well $V_{\rm c}$, equation (\ref{eqVc}), as in neutral Ne@C$_{60}$. Thus, the resonances
 will nearly be $z$-independent. We term such confinement resonances as \textit{conventional confinement resonances}.

To conclude, the above results reveal, and the given explanation proves, that \textit{conventional} confinement resonances in photoionization
cross sections of inner, intermediate and outer subshells of an atom $A$ in $A$@C$_{60}^{z-}$ appear to be almost $z$-independent. This is an interesting general phenomenon.
Another important observation (see figures~\ref{Fig1} and \ref{Fig2}) is that conventional confinement resonances vanish quite rapidly with increasing photon (or photoelectron) energy.
This is in line with a theory of scattering of particles off a potential well/barrier. Indeed, starting at a sufficiently high energy of the outgoing photoelectron,
the coefficient of reflection of the latter off a finite potential well/barrier decreases with increasing energy of the electron. As a result, the interference effect
between the outgoing and scattered off the potential well/barrier photoelectron waves weakens, with increasing energy of the electron, and so are the
associated conventional confinement resonances.  In further, we will term this reasoning as ``conventional thinking''.

\subsection{Coulomb confinement resonances}

Above, trends in the Ne $nl$ photoionization cross sections of Ne@C$_{60}^{z-}$ were considered at photon energies $\omega$ beyond the corresponding $I_{nl}^{z=0}$ thresholds
 of Ne in neutral Ne@C$_{60}$. We now turn the attention to lower photon energy ($\omega < I_{nl}^{z=0}$) parts of figures~\ref{Fig1} and~\ref{Fig2} (to the left of
 the $z=0$ marked line). There, the additional resonance
in each of $\sigma_{\rm 2p}$, $\sigma_{\rm 2s}$ and $\sigma_{\rm 1s}$ of Ne@C$_{60}^{z-}$ is seen to emerge.
It owes its existence to the Coulomb potential $V_{z}$, equation~(\ref{eqVq}), of the charged carbon cage. The $V_{z}$ potential brings up the Coulomb potential barrier
at the outer surface of C$_{60}$. This engenders reflection of the low-energy continuum photoelectron wave
from the Coulomb barrier causing additional resonances one of which is depicted in figures~\ref{Fig1} and \ref{Fig2} at photon energies under discussion. Originally, the emergence of this kind of
a resonance was noted in the Ne ${\rm 1s}$ photoionization of Ne@C$_{60}^{z-}$~\cite{Ne@C60z}, where it was named \textit{Coulomb confinement resonance},
 in view of its association with the Coulomb potential barrier of the charged carbon cage. The present paper establishes that the phenomenon emerges in the photoionization of intermediate
 and outer subshells of the encapsulated atom as well, as a general occurrence.
Coulomb confinement resonances appear to be $z$-dependent, as is clearly exemplified by depicted in figure~\ref{Fig1} $\sigma_{\rm 2p}$'s.
This is because, in this instance,  the energy $\epsilon$ of the outgoing photoelectron
is near or, generally, below the top of the Coulomb potential barrier (see, as illustration, the energy line $\epsilon =0.6$ eV in figure~\ref{Fig3}). This makes the photoionization process to be sensitive to details of the latter, and, hence, to a charged state of the carbon cage as well.

To conclude, the established co-existence of $z$-dependent Coulomb and $z$-independent conventional confinement resonances in photoionization spectra of endohedral \textit{anions}
is an exclusive feature of these systems.

\subsection{Correlation confinement resonances: the resurrection of confinement resonances far above thresholds}

\begin{figure}[h]
\includegraphics[width=16pc]{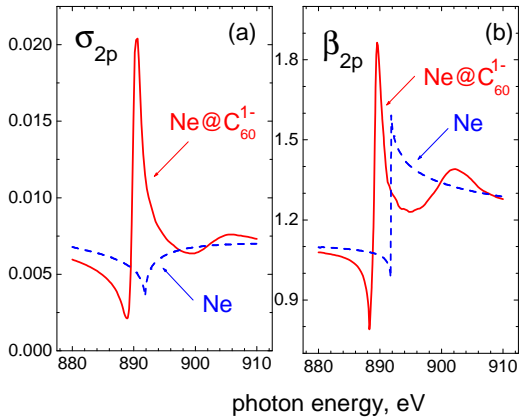}\hspace{2pc}%
\begin{minipage}[b]{14pc}\caption{\label{Fig4} (a): RPAE calculated data for $\sigma_{\rm 2p}(\omega)$ (Mb) of Ne in Ne@C$_{60}^{1-}$  and free Ne
near their ${\rm 1s}$ thresholds, respectively. RPAE calculations included inter-channel coupling between transitions from the ${\rm 1s}$, ${\rm 2s}$
and ${\rm 2p}$ subshells.
(b): The same as in (a) but for the dipole photoelectron angular-asymmetry parameter $\beta_{\rm 2p}(\omega)$.}
\end{minipage}
\end{figure}

In the above, the discussion was related to RPAE calculated data for photoionization cross sections  $\sigma_{nl}(\omega)$ of the encapsulated
Ne atom which were obtained at an intra-shell
interchannel coupling approximation level. However~\cite{DolmAQC09,PranawaXe09,DolmXe08}, the effect of \textit{inter-shell}
interchannel coupling in the encapsulated atom may result in the emergence of new confinement resonances,
termed \textit{correlation confinement resonances}.  These resonances were previously interpreted as resonances
which are induced in an outer-shell photoionization spectrum of
the encapsulated atom by \textit{conventional confinement resonances} in inner-shell photoionization transitions in the atom, via interchannel
coupling. The earlier finding was illustrated by RPAE~\cite{DolmAQC09,DolmXe08} and recently seconded by relativistic RRPA~\cite{PranawaXe09}
calculated data for the Xe ${\rm 5s}$ photoionization of Xe@C$_{60}$ where interchannel coupling between the ${\rm 5s}$ and ${\rm 4d}$
transitions was accounted for. However, the same effect may occur via interchannel coupling with
Coulomb confinement resonances as well. It may even be bigger in this case since Coulomb confinement resonances dominate over conventional confinement resonances;
see figure~\ref{Fig2} for the most illustrative supporting evidence. Furthermore, the effect of interchannel
coupling may show up strongly in an outer-shell photoionization spectrum at photoelectron energies which are thousands eV above the threshold,
when interchannel coupling involves very deep inner-shell transitions. The effect is going to be strong, because
at such big differences in ionization thresholds of the inner and outer subshells, photoionization transitions from the former will be strong whereas those from
the latter will be week, at photon energies above the inner shell threshold.   As a result, both inner-shell Coulomb and conventional confinement resonances may be
effectively ``funneled'' through a thousands-eV-distance to the outer-shell spectrum. However, to which extent the ``funneled'' confinement resonances may indeed perturb
the outer-shell spectrum is not clear.
To clarify this point, we performed RPAE calculations both of the ${\rm 2p}$ photoionization cross section $\sigma_{\rm 2p}$
and dipole photoelectron angular-asymmetry parameter $\beta_{\rm 2p}$
for Ne in Ne@C$_{60}^{1-}$, above the Ne ${\rm 1s}$ ionization threshold. This time, inter-shell interchannel coupling between the
${\rm 1s}$, ${\rm 2s}$ and ${\rm 2p}$ transitions
was included in the calculations. Thus obtained RPAE calculated data for $\sigma_{\rm 2p}$ and $\beta_{\rm 2p}$  for the encaged Ne are depicted in figure~\ref{Fig4} along with
data for free Ne.

One can see that both $\sigma_{\rm 2p}$ and  $\beta_{\rm 2p}(\omega)$ for the encapsulated Ne atom possess a strong sharp resonance at about $890$ eV which is followed
by a lower but broader resonance at about $905$ eV. As a result, these ``encapsulated'' $\sigma_{\rm 2p}$ and  $\beta_{\rm 2p}(\omega)$
differ considerably from  the free Ne $\sigma_{\rm 2p}$ and  $\beta_{\rm 2p}(\omega)$, far above threshold.
Thus, the confinement matters in this case, so that the two prominent resonances
in ``encapsulated'' $\sigma_{\rm 2p}$  and  $\beta_{\rm 2p}(\omega)$ \textit{are} confinement resonances.

The striking novelty of the above finding is that
the resonances emerge far-far above where conventional thinking said they would exist. Indeed, when considering confinement resonances, one normally thinks in terms
of conventional confinement resonances which occur due to the interference between the directly outgoing and reflected off the confining potential photoelectron waves. However,
in line with ``conventional thinking'', as was discussed above, conventional confinement resonances fade away relative rapidly
with increasing energy and do vanish far above threshold. The discovered emergence, or better say resurrection of  confinement resonances in the $\rm 2p$ photoionization spectrum
of Ne@C$_{60}^{z-}$ far above threshold
implies that, in contrast to ``conventional thinking'',  a few eV deep/high confining potential well/barrier may, once again, be felt
by a far-above-potential-barrier-electron. The effect may as well be called
\textit{reemerging confinement effect} for a high-energy scattering electron, as a general phenomenon. This general phenomenon may result in the emergence of far above threshold confinement resonances
in the $nl$ photoionization spectrum of a confined atom, as in the above discussed particular example of the Ne $\rm 2p$ photoionization of Ne@C$_{60}^{1-}$. The latter effect may rightly be termed as
\textit{resurrection of confinement resonances} effect.  Both the reemerging confinement and resurrection of confinement resonances effects
owe their existence to inter-shell interchannel coupling in the encapsulated multielectron atom. Indeed, a trial calculation for Ne@C$_{60}^{1-}$ showed that removal of
the Ne ${\rm 1s}$ transition (and, thus, associated with it Coulomb and conventional confinement resonances)
from RPAE calculations of the Ne ${\rm 2p}$ photoionization leaves no traces of the two resonances in ``encapsulated''
$\sigma_{\rm 2p}$  and  $\beta_{\rm 2p}(\omega)$. As a result,
the $\rm 2p$ photoionization spectra of the confined and free Ne atoms become virtually identical far above threshold, as they previously have been thought
to remain nearly identical at all high energies, on the basis of ``conventional thinking''.
Hence, the resurrected confinement resonances in ``encapsulated''  $\sigma_{\rm 2p}$  and  $\beta_{\rm 2p}(\omega)$, far above threshold, are due to interchannel coupling with
the conventional and Coulomb confinement resonances in the $\rm 1s$ spectrum. The latter are ``funneled'' to the $\rm 2p$ spectrum via interchannel coupling.
 This appears to perturb the outer-shell spectrum of the confined atom dramatically.

Clearly, there is nothing particularly special about the Ne@C$_{60}^{z-}$ system. Therefore, both the reemerging confinement effect and the resurrection of confinement resonances effect
are expected to appear in, and be qualitatively similar for, spectra of other endohedrally confined atoms as well. In other words, the two discovered effects step in as novel general features
of spectra of endohedrally confined atoms $A@C_{60}^{z-}$ the existence and significance of which has been convincingly proven in the performed study.

In conclusion, neither Coulomb or conventional confinement resonances, not to mention the just discovered resurrected confinement resonances far above threshold, have been experimentally
observed yet for technical reasons. We hope that the data presented herein will prompt experimentallists to look into the matter, thereby promoting such
developments.

\ack
This work was supported by the NSF Grant No.\ PHY-$0652704$.

\end{document}